\pdfoutput=1

\documentclass[11pt,twoside,a4paper,cmspaper,final,collab]{cms-tdr}
\def\svnVersion{177019}\def\cmsCernNo{2012-272}\def\cmsCernDate{\today}\def\cmsMessage{Submitted to Physics Letters B}

\begin{document}\cmsNoteHeader{SUS-12-014}

\hyphenation{had-ron-i-za-tion}
\hyphenation{cal-or-i-me-ter}
\hyphenation{de-vices}

\RCS$Revision: 163943 $
\RCS$HeadURL: svn+ssh://svn.cern.ch/reps/tdr2/papers/SUS-12-014/trunk/SUS-12-014.tex $
\RCS$Id: SUS-12-014.tex 163943 2013-01-09 16:02:50Z jhirsch $
\newlength\cmsFigWidth
\ifthenelse{\boolean{cms@external}}{\setlength\cmsFigWidth{0.9\columnwidth}}{\setlength\cmsFigWidth{0.6\textwidth}}
\ifthenelse{\boolean{cms@external}}{\providecommand{\cmsLeft}{top}}{\providecommand{\cmsLeft}{left}}
\ifthenelse{\boolean{cms@external}}{\providecommand{\cmsRight}{bottom}}{\providecommand{\cmsRight}{right}}
\newcommand {\met}{\MET}
\newcommand{\st}{\ensuremath{S_\mathrm{T}}\xspace}
\newcommand {\sq}{\sQua}
\newcommand {\gluino}{\sGlu}

\newcommand {\asq}{\ensuremath{\sq^*}}
\newcommand {\msq}{\ensuremath{M_{\sq}}}

\newcommand {\stmin}{\ensuremath{S^\text{min}_\mathrm{T}}}
\renewcommand{\PYTHIA} {{\textsc{pythia}}\xspace}
\newcommand {\obslim}{1430}
\newcommand {\explim}{1420}

\cmsNoteHeader{SUS-12-014} % This is over-written in the CMS environment: useful as preprint no. for export versions
\title{Search for supersymmetry in events with photons and low missing transverse energy in pp collisions at $\sqrt{s}=7$\TeV}

\date{\today}

\abstract{
Many models of new physics, including versions of supersymmetry (SUSY), predict
production of events with low missing transverse energy, electroweak gauge bosons,
and many energetic final-state particles.  The stealth SUSY model yields this
signature while conserving $R$-parity by means of a new hidden sector in
which SUSY is approximately conserved.  The results
of a general search for new physics, with no requirement on missing
transverse energy, in events with two photons and four or more hadronic
jets are reported.  The study is based on a sample of proton-proton collisions
at $\sqrt{s}=7$\TeV corresponding to 4.96\fbinv of integrated luminosity
collected with the CMS detector in 2011.  Based on good agreement between the
data and the standard model expectation, the data are used to determine
model-independent cross-section limits and a limit on the squark mass in the
framework of stealth SUSY.  With this first study of its kind, squark masses less
than 1430\GeV are excluded at the 95\% confidence level.}

\hypersetup{%
pdfauthor={CMS Collaboration},%
pdftitle={Search for supersymmetry in events with photons and low missing transverse energy in pp collisions at sqrt(s) = 7 TeV},%
pdfsubject={CMS},%
pdfkeywords={CMS, physics}}

\maketitle %maketitle comes after all the front information has been supplied

Models of supersymmetry ~\cite{susy, martin} (SUSY) that conserve
$R$-parity and include a neutral, weakly interacting lightest supersymmetric
particle (LSP) garner much interest because they simultaneously
solve the hierarchy problem, allow unification of the fundamental
interactions, and provide a candidate for dark matter.
Many searches for SUSY focus on the presence of large
missing transverse energy (\met) carried away by the LSP. 
It has been pointed
out~\cite{wacker, lisanti, stealthSusy1}
that this approach neglects well motivated SUSY models that predict low
\met\ without special tuning of masses -- models characterized by
$R$-parity violation~\cite{rpv}, gauge mediated SUSY
breaking~\cite{giudice}, compressed spectra~\cite{martin1, martin2},
and hidden valleys~\cite{strassler}.   As the parameter
space available for high-\met SUSY is reduced by recent results
from the CERN Large Hadron Collider (LHC), low-\met\ alternatives
become more important to study.

Without the discriminating power of large \met, new techniques are required
to reduce standard model (SM) backgrounds.  In general, searches in final states
with many energetic particles, including photons or leptons, are sensitive
to a wide range of new physics models producing heavy particles
with cascade decays to hadronic jets and electroweak gauge bosons.  This range
includes theories of extra dimensions~\cite{appel}, heavy-flavor compositeness~\cite{gregoire}, little
Higgs~\cite{AH, katz}, and even SUSY.  In particular, such searches are
sensitive to the $R$-parity conserving stealth SUSY model~\cite{stealthSusy1, stealthSusy2}, which, as
formulated in this study, produces events with low \met, two photons, and
more than five jets.

The simplest stealth SUSY models assume low scale SUSY breaking
and introduce a new hidden sector of particles at the weak scale in
which only a small amount of SUSY breaking occurs through
interactions with SM fields, resulting in a hidden
sector that is approximately supersymmetric and nearly mass degenerate
hidden sector superpartners.
In this framework, the standard LSP takes on a new role as the lightest
``visible sector'' SUSY particle (LVSP) which decays without violating
$R$-parity into a lighter hidden sector SUSY particle.   In the
subsequent decay of this particle to its SM partner and the true LSP, the near
mass degeneracy leaves little phase space for the true LSP to carry momentum.
In this way, stealth SUSY models naturally produce signatures of low \met
without special tuning of masses.

In this Letter we describe a search for new phenomena (NP) in events with at least
two photons and four or more jets produced in proton-proton collisions at $\sqrt{s}=7$\TeV.
The analysis is based on a data sample corresponding to $4.96\pm0.11\fbinv$ of
integrated luminosity~\cite{lumi} collected in 2011 with the Compact Muon Solenoid (CMS) detector
at the LHC.  We perform a model-independent search for an excess of events with
respect to the SM expectation, and we interpret this search in the framework of stealth SUSY.
To minimize the impact of SM backgrounds on the sensitivity to NP, event excesses are determined as
functions of final-state jet multiplicity and the scalar sum of the transverse momentum
\pt of jets, photons, and \met (for \met$>20$\GeV) in each event; this sum is
referred to as \st.  We define \met as the magnitude of the negative of the vector 
sum of the \pt of all final state objects in the event; the definitions of other objects 
included in the sum are described below.  

\begin{figure}[hbtp]
\begin{center}
\includegraphics[width=\cmsFigWidth]{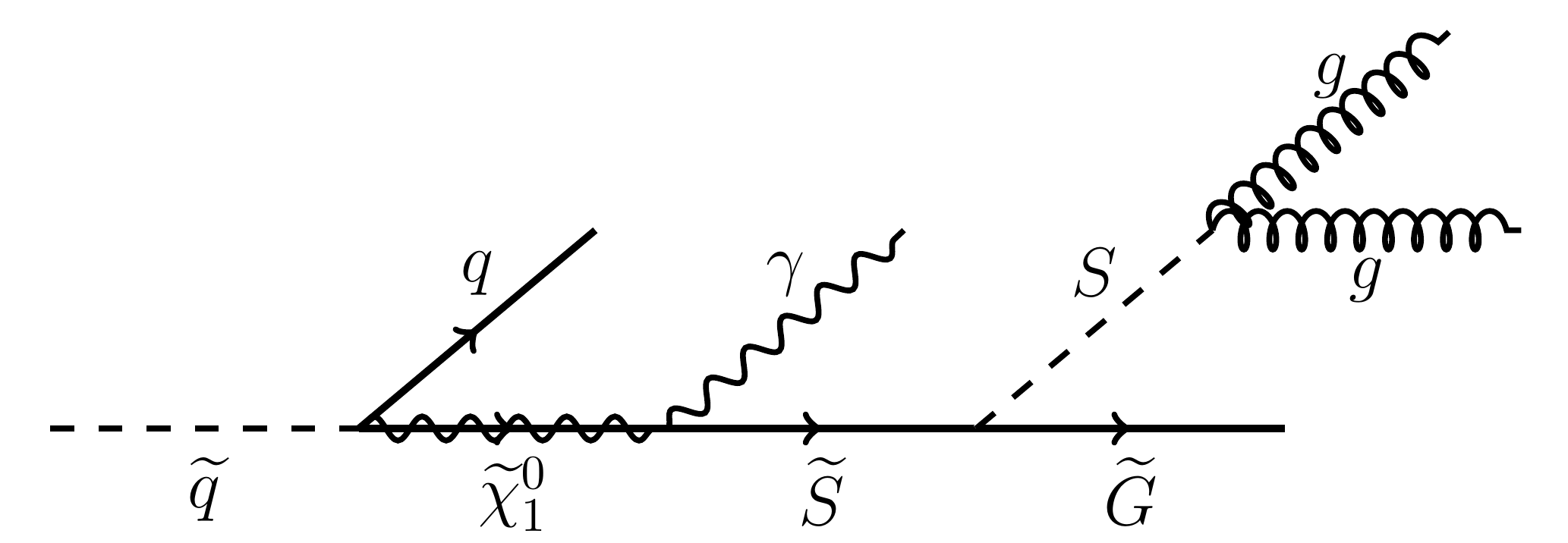}
\caption{Decay of a squark in stealth SUSY.  See text for a complete description.}
\label{fig:squarkDecay}
\end{center}
\end{figure}

We consider a model that includes degenerate light squarks (\sq),
a ``bino-like'' LVSP $\widetilde{\chi}_{1}^{0}$,
a gluino (\gluino) with mass of 1500\GeV, and
a hidden sector containing a singlet state S and its fermionic ``singlino''
superpartner $\widetilde{\textrm{S}}$.   The model is similar to squark-antisquark (\sq\asq)
production with the decay $\widetilde{\textrm{q}}\rightarrow \textrm{q}\widetilde{\chi}_{1}^{0}$
described in Ref.~\cite{sms} with two differences: the addition of the hidden sector and
the participation of \gluino\ in the production mechanism.
After initial production of two $\widetilde{\chi}_{1}^{0}$ and jets, each $\widetilde{\chi}_{1}^{0}$
decays into the hidden sector producing a photon and $\widetilde{\textrm{S}}$, which subsequently
decays to S and a gravitino, $\widetilde{\textrm{S}} \rightarrow \mathrm{S}\widetilde{\textrm{G}}$.
The S state is even under $R$-parity and decays to jets via S$\rightarrow$gg.
The resulting $\widetilde{\textrm{G}}$ LSP has small momentum because the hidden sector superpartners (S, $\widetilde{\textrm{S}}$)
are nearly mass degenerate,
so the final state tends to have low \met.  The decay of a squark in this model
is shown in Fig.~\ref{fig:squarkDecay}.

The model is characterized by the masses of the particles in the decay chain.  We
consider a range of squark masses \msq\ from
400 to 2000\GeV in steps of 100\GeV.  We make the following assumptions inspired by
benchmark points described in Ref.~\cite{stealthSusy1}: $\widetilde{\textrm{S}}$ and S
masses of 100\GeV and 90\GeV, respectively; $\widetilde{\chi}_{1}^{0}$ with mass equal
to $\frac{1}{2}\msq$; and branching fractions of unity for the decays of
$\widetilde{\chi}_{1}^{0}$, $\widetilde{\textrm{S}}$, and S described above.
The production cross section for this process (and its uncertainty) is
calculated as a function of \msq\ at next-to-leading order (NLO) accuracy
including the resummation of soft gluon emission at next-to-leading logarithmic (NLL)
accuracy as described in Ref.~\cite{Beenakker:1996ch, Kulesza:2008jb, Kulesza:2009kq,
Beenakker:2009ha, Beenakker:2011fu, Kramer:2012bx}.

A detailed description of the CMS detector can be found in Ref.~\cite{cmsdet}.
The CMS coordinate system is right-handed with the origin at the center of the
detector, the $x$-axis directed toward the center of the LHC ring, and the $y$-axis
directed upward;  $\phi$ is the azimuthal angle, $\theta$ is the polar angle, and
$\eta = -\ln[\tan(\theta/2)]$ is the pseudorapidity.  The central feature of the
CMS apparatus is a superconducting solenoid of 6\unit{m} inner diameter that surrounds
a silicon pixel and strip tracker, covering the region $|\eta| < 2.5$, as well
as a lead-tungstate crystal electromagnetic calorimeter (ECAL) and a brass/scintillator
hadron calorimeter (HCAL), both covering $|\eta| < 3$.  Muons are detected by
gas-ionization detectors embedded in the steel flux return yoke covering the
range $|\eta| < 2.4$.

The triggers, event reconstruction methods, and selection criteria for photons and jets
are identical to those used in the CMS search for SUSY in events with
photons, jets, and \met~\cite{pasSUS12001}.
Events are recorded with the CMS two-level trigger system requiring the presence
of one photon with transverse energy ($\ET$) greater than 36\GeV and a second with
$\ET>22$~\GeV.  To suppress jets giving rise to photon candidates,
these triggers require the latter to be isolated from other activity in the tracker,
ECAL, and HCAL.  As instantaneous luminosity increased throughout 2011,
isolation requirements were gradually changed to keep the trigger rate approximately constant, but
the isolation in the trigger is always less restrictive than offline requirements described below.

Photon candidates are reconstructed from clusters of energy in the
ECAL barrel with $\vert\eta\vert<1.44$.  Candidate events are
required to have a leading photon with $\ET>40$\GeV and an
additional photon with $\ET>25$\GeV; at these
thresholds the triggers are more than 99\% efficient.
We require the ECAL cluster shape to be consistent with that
expected for photons, and the energy detected in HCAL
in the direction of the photon shower to not exceed 5\% of
the ECAL energy.  We ensure isolation from other activity in the event by requiring that
the scalar $\ET$ sum of tracks and calorimeter deposits within
$\Delta R = \sqrt{ (\Delta\eta)^2+(\Delta\phi)^2} = 0.3$ of the photon candidate's
direction be less than 6\GeV after correcting for contributions from the products
of additional collisions in the event (pile-up) and the candidate itself.  These
criteria efficiently select both photons and electrons; candidates that cannot be
matched to hit patterns in the pixel detector are considered photons.

Jets are reconstructed with the particle-flow algorithm
\cite{PFT-09-001}, which simultaneously reconstructs all
particles produced in a collision based on information from
all detector subsystems and identifies each as a charged or
neutral hadron, photon, muon, or electron.   All of these particles
are clustered into jets with the anti-$k_{\rm T}$ clustering algorithm~\cite{antikt}
with radius parameter of 0.5.   To remove jets arising from
potential instrumental and non-collision backgrounds, we require
the fraction of jet energy coming from charged and neutral electromagnetic
deposits to be less than $0.99$, the neutral hadron fraction to be
less than $0.99$, and the charged hadron fraction to be greater
than zero.   The jet energy and momentum are corrected for the nonlinear
response of the calorimeter and the effects of pile-up.
Jets are required to have $\pt>20$~\GeV, $|\eta|<2.4$, and
to be isolated from photon candidates by $\Delta R>0.5$.

We determine the fraction of stealth SUSY events that
pass these reconstruction procedures and selection criteria
(the ``acceptance'') and the leading order plus leading
logarithm (LO) cross section for stealth SUSY using the
\PYTHIA 6.424 event generator~\cite{Sjostrand:2006za},
the D6T underlying event tune~\cite{d6t}, the CTEQ6L1~\cite{cteq}
parton distribution functions (PDFs), and a full
simulation of the CMS detector based on {\sc GEANT4}~\cite{Agostinelli:2002hh}.
The acceptance when including a requirement for
five or more jets rises monotonically from 17\% to 35\%
for \msq\ from 400 to 1100\GeV and falls monotonically from
35\% to 25\% for \msq\ from 1100 to 2000\GeV as additional
hadronic activity makes it less likely that photons will be isolated.
The acceptance when requiring exactly four jets
is 12--13\% (1--2\%) of the $\geq$5-jet acceptance
for $\msq\leq 1500$\GeV ($\msq>1500$\GeV).  The LO acceptance for each
of the dominant subprocesses (pp $\rightarrow$ \sq\sq, \sq\asq, \sq\gluino, and \gluino\gluino)
is weighted with the appropriate NLO/LO $K$-factor;
the corrected acceptance is within $5\%$ of the LO acceptance at
all \msq.

The SM events satisfying the selection criteria come mainly from direct 
production of two photons or a single photon and a jet, which is misidentified 
as a photon.  In both cases, additional jets come from radiation via quantum 
chromodynamics.  These events are divided into three samples: a signal-rich ``search sample''
comprising events with four or more jets and $\st>700$\GeV, a signal-depleted ``jet
multiplicity sideband'' (JMSB) composed of events with two or three
jets and $\st>600$\GeV, and a signal-depleted ``$\st$ sideband''
composed of events with four or more jets and
$600\GeV<\st<700~\GeV$.
We examine the search sample for evidence of NP in the form of an
excess of events  over the SM expectation.   

The SM expectation is estimated from the data based on the observation that the shape of the 
$\st$ spectrum of the SM background is independent of jet multiplicity. 
This multiplicity invariance arises because the $\st$ of an event is dominated by the initial
hard parton-parton scattering process; additional radiation, which is largely 
collinear with incoming or outgoing partons, does not have a large effect 
on event $\st$.  For this reason, we are able to take the $\st$ shape from 
the JMSB and the normalization from the $\st$ sideband.

This method of background estimation was first used in the CMS search for 
black holes~\cite{cmsBH2010, cmsBH}, in which the jet-multiplicity invariance of the
\st shape in jet dominated events was demonstrated.  We confirm
that this invariance holds for events with photons in addition to jets
using a data sample of events with one photon and two or more jets (photon+jets).
The event selection criteria for this sample are the same as those described above
except for changes required by differences in the trigger and event topology:
we require that events include a single photon with $\ET>$ 80\GeV and
$H_\mathrm{T}>$ 450\GeV, where $H_\mathrm{T}$ is defined as the scalar sum of the \pt of
jets with $\pt>$ 40\GeV and $\vert\eta\vert<3.0$.

\begin{figure}[hbtp!]
\begin{center}
\includegraphics[width=0.45\textwidth]{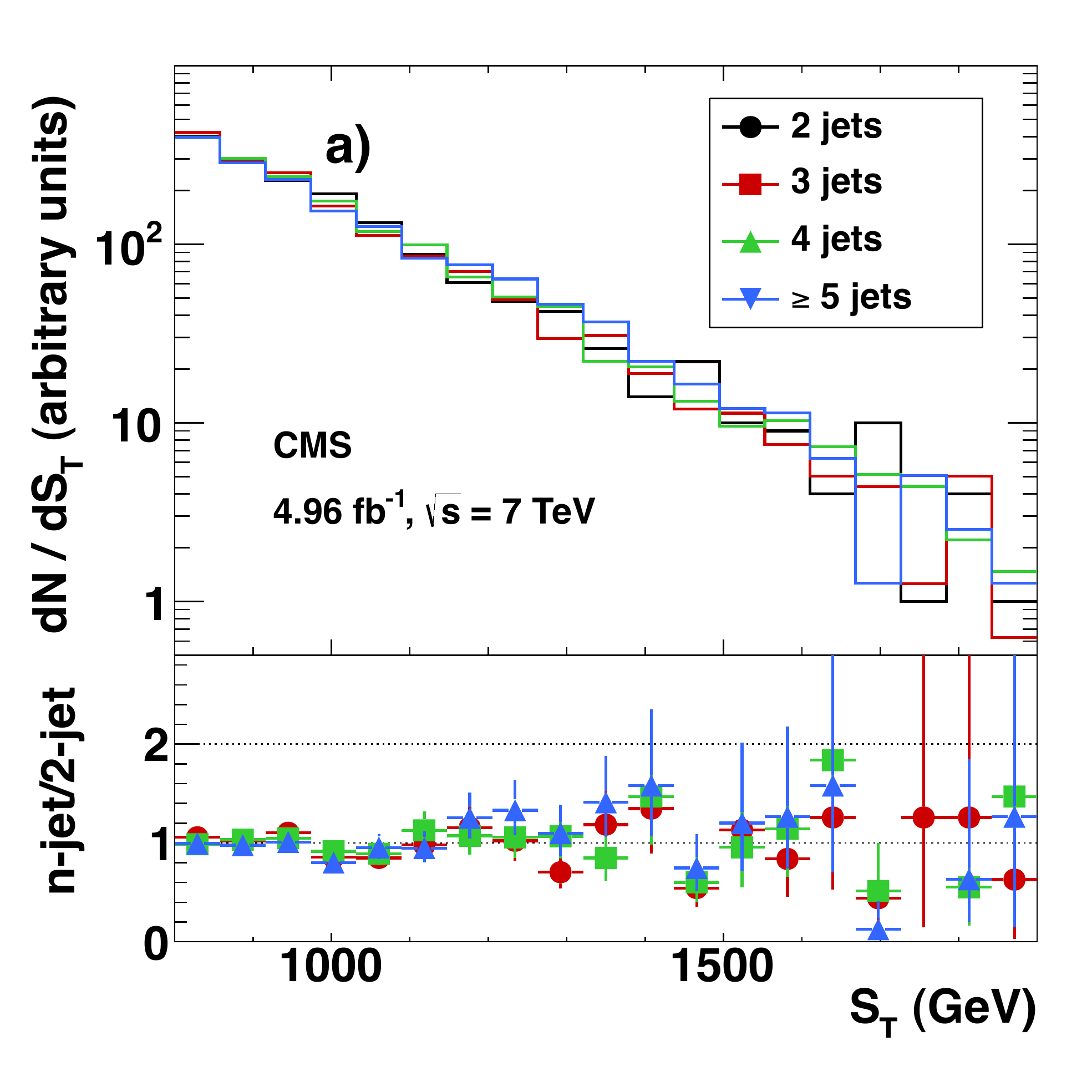}
\includegraphics[width=0.45\textwidth]{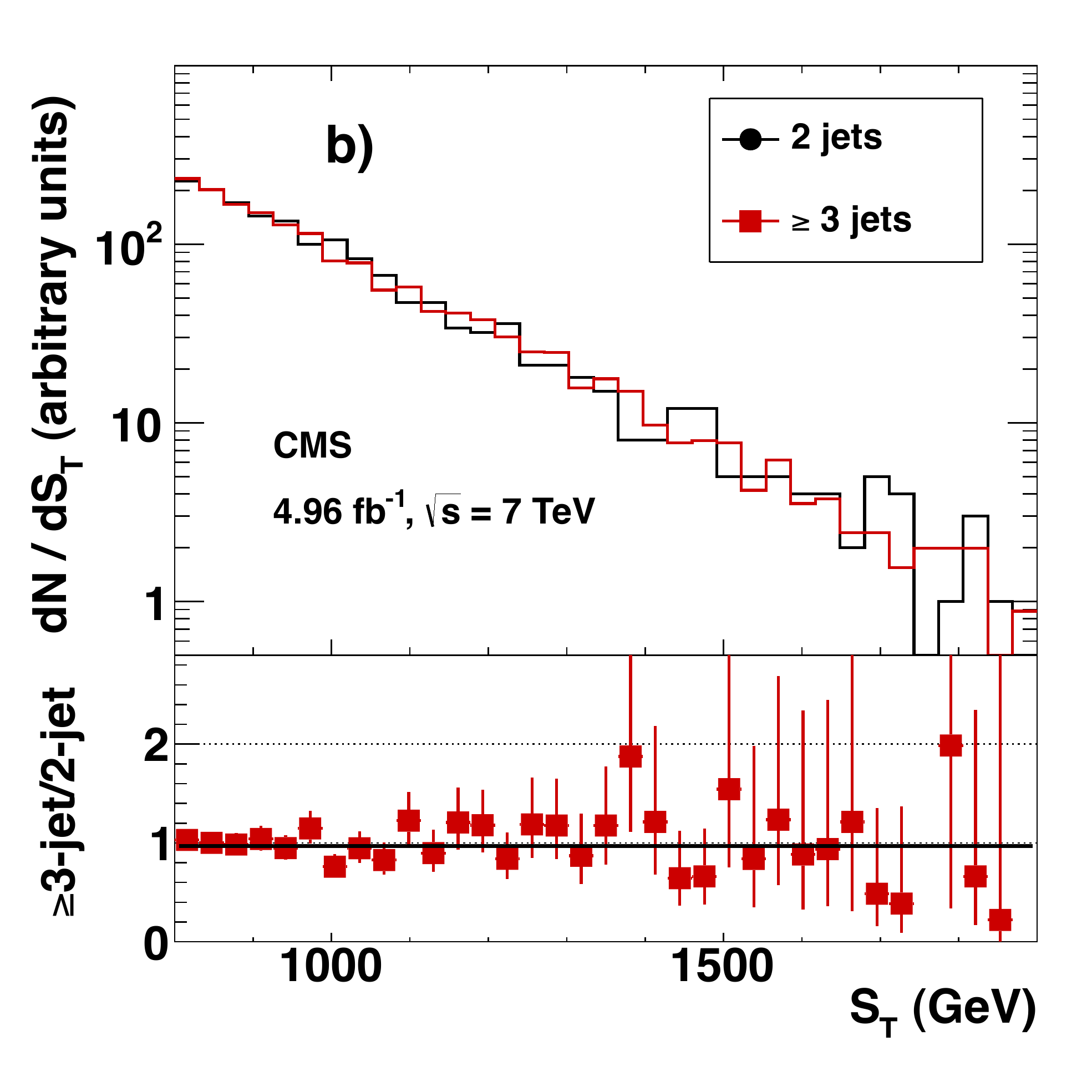}
\caption{\st spectra from the photon+jets data control sample,
area-normalized for $\st>800$\GeV.  We show (a) the spectra for events
with two, three, four, and five or more jets along with the $n$-jet/$2$-jet ratio
where $n$ =  3, 4, or $\geq$5; and (b) the spectra for events with
two and three or more jets along with the $\geq$3-jet/$2$-jet ratio.}
\label{fig:st}
\end{center}
\end{figure}

In Fig.~\ref{fig:st} we compare the \st spectra for five subsamples of this
photon+jets dataset characterized by jet multiplicity: the 2-jet, 3-jet,
4-jet, $\geq$5-jet, and $\geq$3-jet samples.  We show these spectra (area-normalized for
$\st>800$\GeV) along with the $n$-jet/$2$-jet ratio where $n$ = 3, 4, $\geq$5,
and $\geq$3.  The shape of each ratio is consistent with a flat line within the statistical
uncertainty.  As an example, the fit of the $\geq$3-jet/$2$-jet ratio from Fig.~\ref{fig:st}b
with the function $a+bx$ where $x\equiv \st/7\TeV$ yields $a=1.0\pm0.2$
and $b=-0.3\pm1.1$ (statistical uncertainty).  A slope of this size would affect the
expected background rate by 6\%, which is negligible compared to the systematic
uncertainty from other sources (described below).

As introduced above, we model the shape of the \st distribution for
the background by fitting the JMSB data with the function $1/x^{\alpha}$, where $x$
is defined above and $\alpha$ is a free parameter.  We find $\alpha=5.02\pm0.32$ from
the fit of the entire JMSB; in the 2-jet (3-jet) subsample of the JMSB we find
a best fit value of $\alpha=5.00\pm0.45$ ($5.03\pm0.45$).
This background shape is normalized for use in the search
sample with data in the \st sideband.

\begin{figure}[hbtp!]
\begin{center}
\includegraphics[width=\cmsFigWidth]{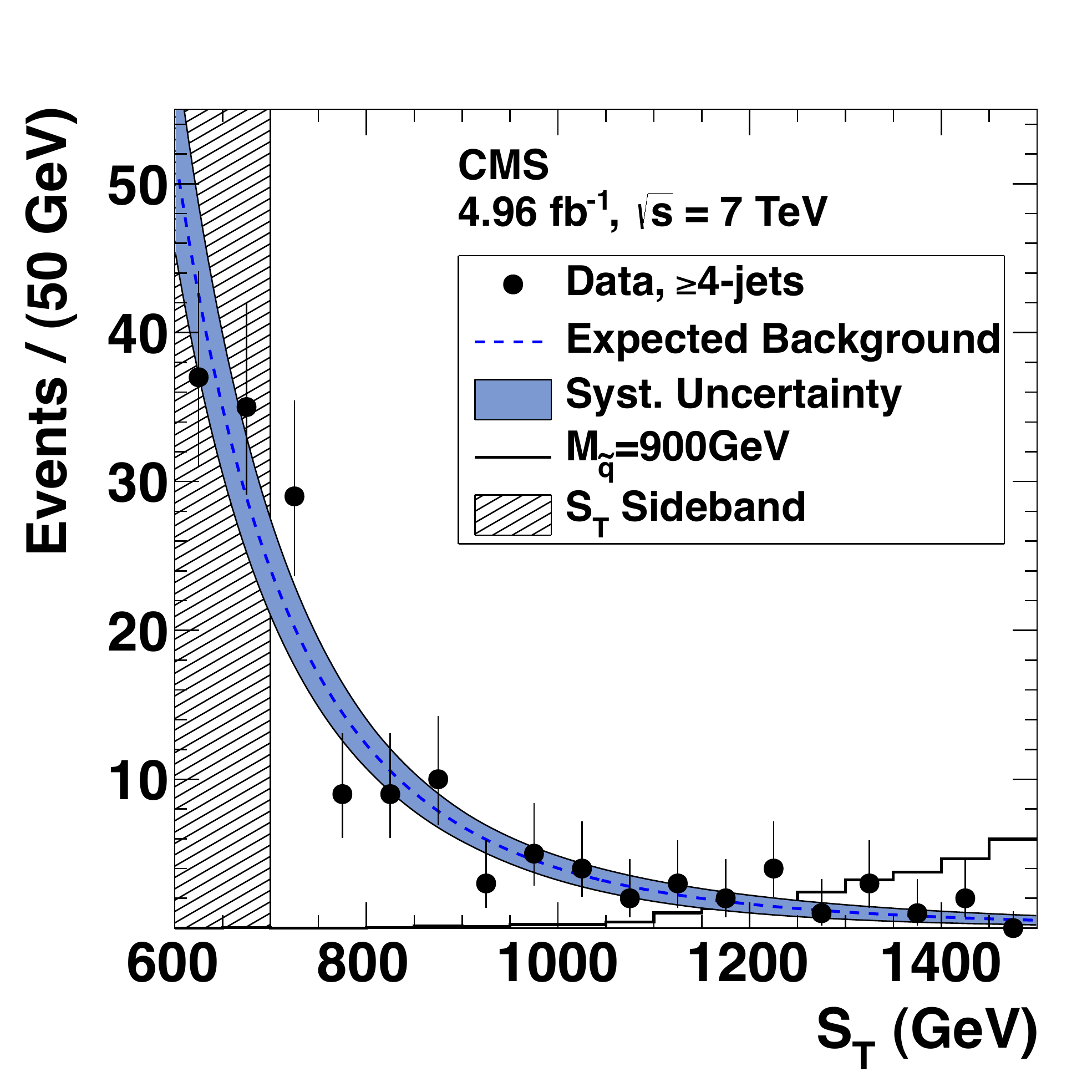}
\caption{Observed \st spectrum, background expectation with systematic uncertainty, and
predicted signal for a squark mass of 900\GeV in events with four or more jets.}
\label{fig:data}
\end{center}
\end{figure}

We compare this background prediction to the observed
\st spectrum in the search sample in Fig.~\ref{fig:data} along
with the systematic uncertainty on the prediction (described below).  The data
are in good agreement with the background expectation.  The probability for the
most significant excess (at $700\GeV<\st<750\GeV$) to have local significance
as high or higher than that observed is 0.08.

Supported by the good agreement of the data with the background expectation,
we use the data to compute $N^\text{lim}_\text{exc}(\stmin)$, the upper limit
on the number of events exceeding the SM expectation as a function of
lower \st threshold.
Limits are computed at the 95\% confidence level (CL) with the modified frequentist
CL$_\mathrm{s}$ method~\cite{cls_read, cls_junk} based on a profile likelihood ratio test
statistic constructed from the Poisson probability for the observed number of events given
the expectations for background and signal.   For the model-independent cross-section
limits, we separately compute $N^\text{lim}_\text{exc}(\stmin)$ for two categories of jet
multiplicity, $\geq$4-jet and $\geq$5-jet.  For stealth SUSY limits, we compute
$N^\text{lim}_\text{exc}(\stmin)$ using exclusive 4-jet and $\geq$5-jet
jet-multiplicity bins combined as described in Ref.~\cite{higgsComb};
the inclusion of the 4-jet bin improves the expected cross-section
limit by 7.5\% at \msq=1400\GeV.

Using the measured integrated luminosity, we convert $N^\text{lim}_\text{exc}(\stmin)$
into a model-independent limit on the product of the acceptance and cross section
of any new process that may be present in addition to the SM (the ``effective NP cross section'').
Using the integrated luminosity and the signal acceptance, we convert
$N^\text{lim}_\text{exc}(\stmin)$ into a limit on the stealth SUSY cross section
as a function of \msq.  For each value of \msq, we
use the value of \stmin\ that maximizes the sensitivity to the stealth SUSY signal
accounting for systematic uncertainty on the background expectation; these
\stmin\ values range from 800 to 2400\GeV for \msq\ from 400 to 2000\GeV.

Systematic uncertainties on the expected numbers of background events, the
signal acceptance, and the integrated luminosity are treated as nuisance parameters
with a Gamma function prior distribution for the uncertainty related to the
normalization of the background prediction and log-normal prior distributions
for the other uncertainties.
The theoretical uncertainty on the predicted cross section does not enter the
cross-section limit, but is used in determining the limit on \msq.

We estimate that the following sources of uncertainty affect our
knowledge of the expected background at the stated level: statistical uncertainty
from the background normalization method (15\%), statistical
uncertainty on the background shape (10--55\%), and the choice of background
function (5--100\%) (ranges indicate dependence on \msq\ and \stmin).  We use
the change in background expectation obtained by varying
the shape parameter by $\pm$1 standard deviation (SD) as the statistical uncertainty on the
background shape.  We estimate the uncertainty related to the choice of
background function by constructing background models from the alternate fit functions
$\mathrm{e}^{\alpha x}$ and $1/x^{\alpha_0+\alpha_1\log(x)}$ instead of the nominal $1/x^{\alpha}$; we take the largest
alternate-nominal difference in the background expectation as the uncertainty.
The relative effect of these background uncertainties is large at high \st
where very few background events are expected.

The following sources of uncertainty affect our knowledge of the expected numbers
of signal events at the stated level: jet
energy scale (3\%), statistical uncertainty on signal acceptance from finite
simulated samples (2\%),  and the measurement of integrated luminosity
(2.2\%)~\cite{lumi}.   The effects on the signal acceptance of variations in
pile-up and PDFs are less than 1\%.  The theoretical uncertainty on the predicted cross
section related to PDFs, renormalization and factorization scales, and $\alpha_S$
variations is estimated to be 9--57\% for \msq\ from 400 to 2000\GeV;
the dominant source is the PDF uncertainty.

\begin{figure}[hbtp]
\begin{center}
\includegraphics[width=0.45\textwidth]{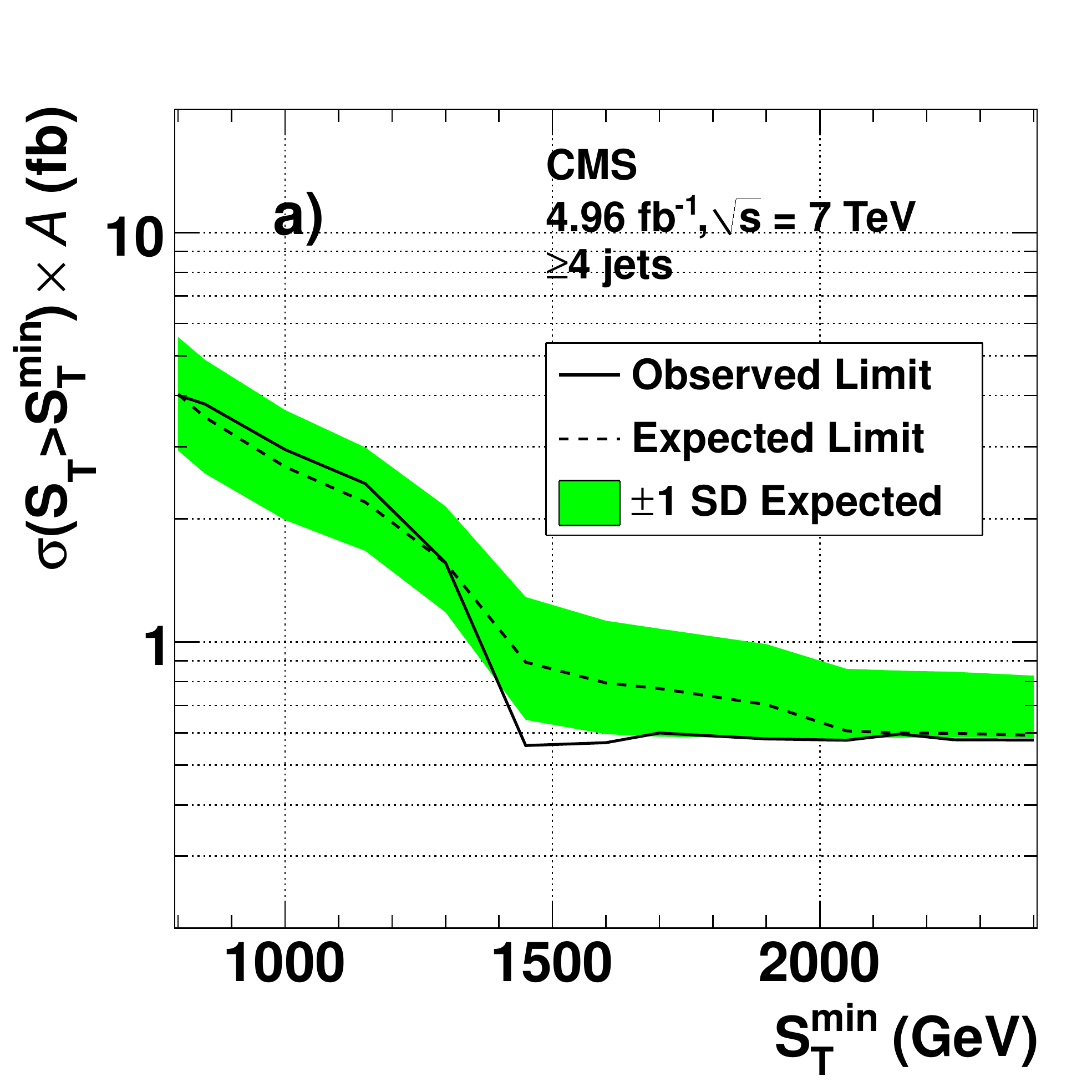}
\includegraphics[width=0.45\textwidth]{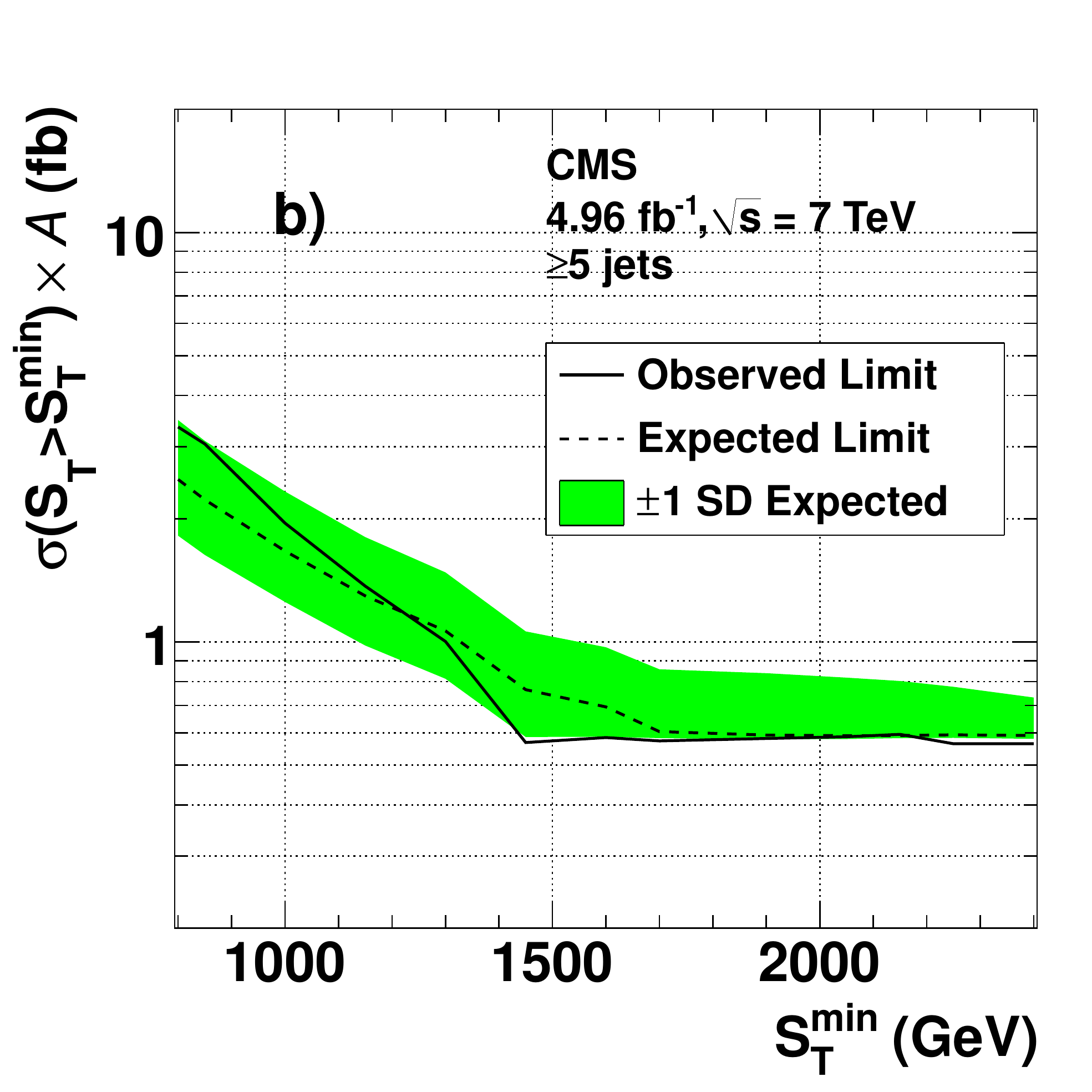}
\caption{Model-independent limit at the 95\% CL on the product of the acceptance and cross section
for events with (a) four or more jets and (b) five or more jets.  We show the observed limits,
median expected limits, and a band corresponding to $\pm1$\unit{SD} on the median expected limits.}
\label{fig:limit_45j}
\end{center}
\end{figure}

\begin{figure}[hbtp]
\begin{center}
\includegraphics[width=0.45\textwidth]{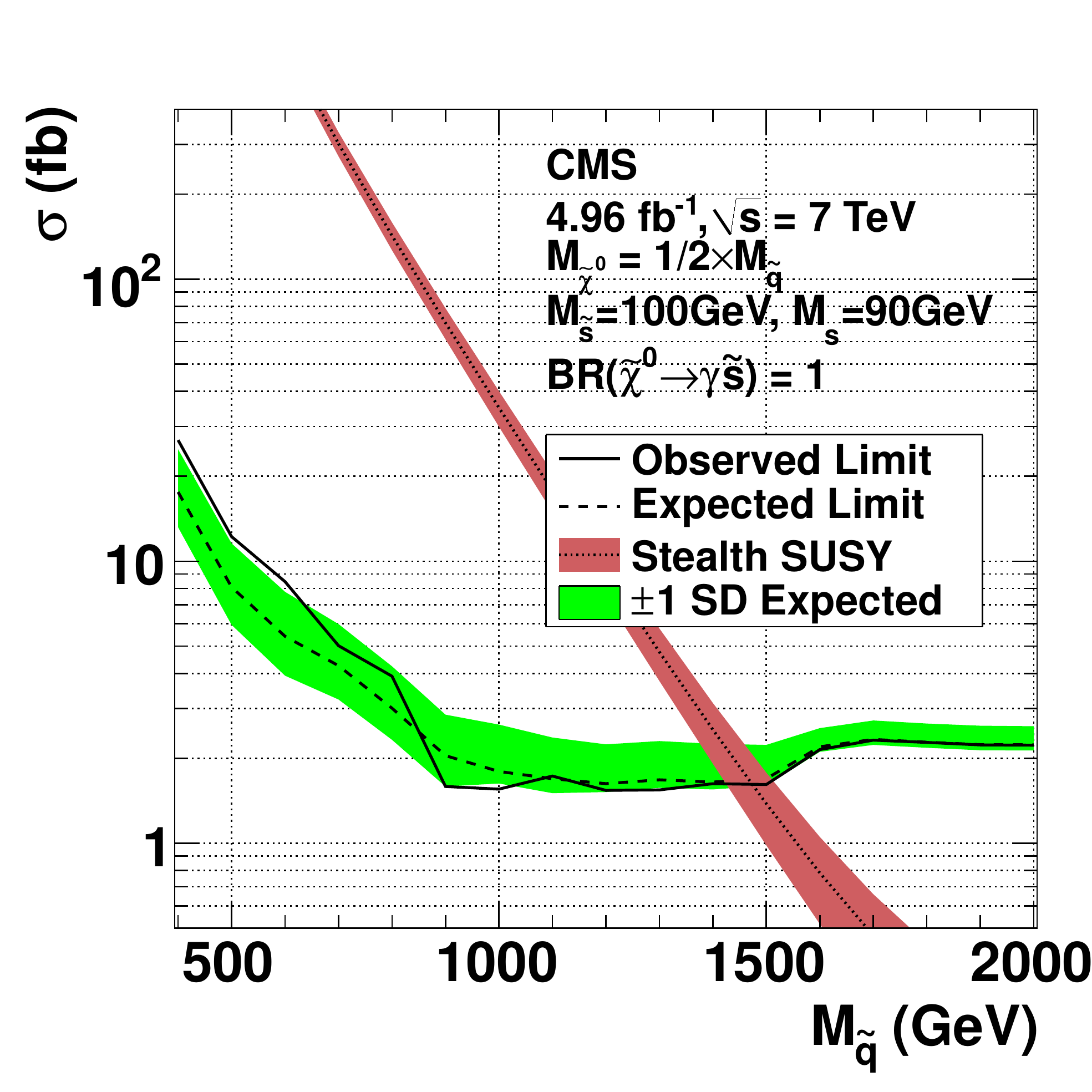}
\caption{Stealth SUSY cross-section limit at the 95\% CL as a function of \msq.  We show the
observed limit, median expected limit, and a band corresponding to ${\pm}1$\unit{SD}
on the median expected limits.  We also show the predicted NLO+NLL cross section from stealth
SUSY with a band denoting ${\pm}1$\unit{SD} from theoretical uncertainty.}
\label{fig:limit}
\end{center}
\end{figure}

Figure~\ref{fig:limit_45j} provides the model-independent limit on the effective NP
cross section as a function of \stmin\ for events with four or more jets and five or
more jets. We show the observed limits, median expected limits, and a band
corresponding to ${\pm}1$\unit{SD} on the median expected limits.
In Fig.~\ref{fig:limit} we show the same curves for the stealth SUSY cross-section limit as
a function of \msq.  We also show the predicted NLO+NLL cross section with a band denoting ${\pm}1$\unit{SD}
from theoretical uncertainty.  Based on the intersection of the cross-section
limit and the $-1$\unit{SD} edge of the predicted cross-section band, we exclude squarks with
\msq$<$\obslim\\GeV for the stealth SUSY model described above.

In summary, we perform a search for NP in events with two photons and
four or more jets.  The selection requirements are general and
provide sensitivity to a broad range of NP phenomena.  We observe no excess over
the SM expectation in a data sample corresponding to 4.96\fbinv of
integrated luminosity collected in 2011.  We determine
model-independent limits on the effective NP cross section of
0.6 to 4\unit{fb} depending on \stmin.  We also compute the limit on the stealth
SUSY cross section as a function of \msq.  Comparing this limit to the predicted 
stealth SUSY cross section, we exclude the production of squarks with 
$\msq<$\obslim\GeV.  Existing SUSY searches based on photons and 
\met~\cite{pasSUS12001} are insensitive to most of the 
stealth SUSY region excluded by this analysis; this region is characterized by low \met
predicting $1.5\pm0.1$ ($21.3\pm1.6$) events with $\met>100$\GeV
for \msq\ of 1400 (800)\GeV.
This is the first limit on the parameters of the stealth SUSY model.

We thank Joshua Ruderman for help in stealth SUSY event generation.  We congratulate our colleagues in the CERN accelerator departments for the excellent performance of the LHC machine. We thank the technical and administrative staff at CERN and other CMS institutes, and acknowledge support from BMWF and FWF (Austria); FNRS and FWO (Belgium); CNPq, CAPES, FAPERJ, and FAPESP (Brazil); MES (Bulgaria); CERN; CAS, MoST, and NSFC (China); COLCIENCIAS (Colombia); MSES (Croatia); RPF (Cyprus); MEYS (Czech Republic); MoER, SF0690030s09 and ERDF (Estonia); Academy of Finland, MEC, and HIP (Finland); CEA and CNRS/IN2P3 (France); BMBF, DFG, and HGF (Germany); GSRT (Greece); OTKA and NKTH (Hungary); DAE and DST (India); IPM (Iran); SFI (Ireland); INFN (Italy); NRF and WCU (Korea); LAS (Lithuania); CINVESTAV, CONACYT, SEP, and UASLP-FAI (Mexico); MSI (New Zealand); PAEC (Pakistan); MSHE and NSC (Poland); FCT (Portugal); JINR (Armenia, Belarus, Georgia, Ukraine, Uzbekistan); MON, RosAtom, RAS and RFBR (Russia); MSTD (Serbia); SEIDI and CPAN (Spain); Swiss Funding Agencies (Switzerland); NSC (Taipei); ThEP, IPST and NECTEC (Thailand); TUBITAK and TAEK (Turkey); NASU (Ukraine); STFC (United Kingdom); DOE and NSF (USA).

\bibliography{auto_generated}   % will be created by the tdr script.

\cleardoublepage \appendix\section{The CMS Collaboration \label{app:collab}}\begin{sloppypar}\hyphenpenalty=5000\widowpenalty=500\clubpenalty=5000\textbf{Yerevan Physics Institute,  Yerevan,  Armenia}\\*[0pt]
S.~Chatrchyan, V.~Khachatryan, A.M.~Sirunyan, A.~Tumasyan
\vskip\cmsinstskip
\textbf{Institut f\"{u}r Hochenergiephysik der OeAW,  Wien,  Austria}\\*[0pt]
W.~Adam, E.~Aguilo, T.~Bergauer, M.~Dragicevic, J.~Er\"{o}, C.~Fabjan\cmsAuthorMark{1}, M.~Friedl, R.~Fr\"{u}hwirth\cmsAuthorMark{1}, V.M.~Ghete, J.~Hammer, N.~H\"{o}rmann, J.~Hrubec, M.~Jeitler\cmsAuthorMark{1}, W.~Kiesenhofer, V.~Kn\"{u}nz, M.~Krammer\cmsAuthorMark{1}, I.~Kr\"{a}tschmer, D.~Liko, I.~Mikulec, M.~Pernicka$^{\textrm{\dag}}$, B.~Rahbaran, C.~Rohringer, H.~Rohringer, R.~Sch\"{o}fbeck, J.~Strauss, A.~Taurok, W.~Waltenberger, G.~Walzel, E.~Widl, C.-E.~Wulz\cmsAuthorMark{1}
% ...
\vskip\cmsinstskip
\dag:~Deceased\\
1:~~Also at Vienna University of Technology, Vienna, Austria\\
% ...

\end{sloppypar}
\end{document}